# Angular harmonics of the excitonic polarization conversions effect

A.V. Koudinov

A.F. Ioffe Physico-Technical Institute of RAS, Politekhnicheskaya 26,
194021 St.-Petersburg, Russia

*We suggest a phenomenological theory of the polarization conversions effect, an excitonic analog of the first-order spatial dispersion phenomena which is, however, observed in the photoluminescence rather than in the passing light. The optical polarization response of a model system of electrically neutral quantum dots subject to the magnetic field along the growth axis was calculated by means of the pseudospin method. All possible forms of the polarization response are determined by nine different field-dependent coefficients which represent, therefore, a natural basis for classification of a variety of conversions. Existing experimental data can be well inscribed in this classification scheme. Predictions were made regarding two effects which have not been addressed experimentally.*

**Introduction**

The term 'polarization conversions' appeared in mid-1990s with regard to the polarization phenomena observed at the conditions of optical pumping of GaAs/(Al,Ga)As layered nanostructures.[1,2] In particular, the external magnetic field applied along the growth axis caused the appearance of the circularly polarized component of the photoluminescence (PL) at a linearly polarized optical excitation or, vice versa, appearance of the linearly polarized component of the PL at a circularly polarized excitation. The 'linear-to-linear conversion', i.e., merely the rotation of the plane of polarization of the PL with respect to that of the excitation light, was also observed.[2] Taking a broader view of things, one can consider as 'conversions' all the collection of the exciton-mediated relationships between the three polarization parameters of the excitation light and the three polarization parameters of the PL. With such an interpretation, well-known effects of optical orientation and optical alignment of excitons[3] are particular forms of the conversions.

Later on, the conversions effect was reported for quantum dot layers.[4,5] Reciprocal conversions of the linear and the circular polarizations were found to occur even without any magnetic field applied.[6,7] The 'swings of optical alignment' were observed.[8] Lately, very efficient polarization conversions by single quantum dots were reported.[9,10] Incidentally, the recent experimental studies[6–10] widely exploited the *method of angular harmonics of polarization*, where the sample was rotated about the growth axis while the polarization responses (various conversions) were studied in their dependency on the rotation angle $\varphi$. This fact motivated us to undertake an explicit analysis of possible $\varphi$-dependences of the polarization responses – in more general terms than it has been attempted in Ref.8.

There exists a profound analogy between the polarization conversions (which are excitonic effects by definition and which are observed in the PL) and a group of first-order spatial dispersion effects like linear or elliptic birefringence and Faraday rotation.[11,12] This analogy stands behind the similarity between the pseudospin method (which was first adopted to describe the problem of polarization conversions by Dzhioev et al.[2] and, later, used in many studies) and the Poincaré sphere construction, which was developed well earlier by Mallard and Poincaré[13] to describe the polarization of light in a birefringent gyrotropic media. We shall use the pseudospin method to calculate various forms of the polarization response (circular and linear PL polarization degrees responding 100% circular or linear optical excitation) as functions of the sample orientation $\varphi$ and the value of magnetic field $B$, with the field **B** directed along the growth axis and the light propagation direction.

**Models and results**

We shall calculate optical polarization responses of a layer of uncharged quantum dots where neutral excitons are created by the polarized optical excitation. We consider two-step models of the exciton evolution, and the relaxation is not taken into account. The exciton is created in the upper (excited) state, subsequently jumps to the lower (ground) state to remain there until the recombination. The lifetime in the upper state, limited by the relaxation to the lower state, is $\tau_{upper}$; the lifetime in the lower state, limited by the recombination, is $\tau_{lower}$. This scheme corresponds to typical experimental conditions and gives a more complete description of optical experiments[2,8,14] than the rudimentary one-step model which is frequently used.

The essence of the conversions phenomenon is in reconstruction of the symmetry of the bright exciton states by the magnetic field $B$ applied along the growth axis. At $B = 0$, the two bright states correspond to two orthogonal linear dipole oscillators and are detuned in the energy scale by a value of anisotropic exchange splitting $\delta_1$. At $B$ strong enough, they correspond to two opposite circular dipole oscillators and are detuned by a larger value $\sqrt{\delta_1^2 + \delta_Z^2}$ where the Zeeman energy $\delta_Z \propto B$. Thus the characteristic field range where the conversions occur is limited by $\delta_Z$ several times $\delta_1$.

We shall use a pseudospin description of the time evolution of the polarization state of the exciton. The pseudospin concept works well as long as only the bright excitons states take part while dark excitons are not involved. The dark states are off the bright states by the value of the isotropic exchange splitting $\delta_0$, and, typically, $\delta_1 \sim 0.1\delta_0$. So in all the essential $B$ range $\delta_Z \sim \delta_1$ the bright states are well isolated, which justifies introduction of the pseudospin.

The components of the of the polarization response of the QD layer to the linearly- or circularly-polarized excitation were sought by conventional means. The Bloch equation for the exciton pseudospin $\vec{S}$

$$\dot{\vec{S}} = \vec{\Omega} \times \vec{S} \tag{1}$$

was solved in the upper and in the lower state, where each state was characterized by its own effective Larmor frequency $\vec{\Omega}$ depending on the exciton g-factor, the applied magnetic field $B$, the value of $\delta_1$ and the in-plane orientation of the QD potential. The time-dependent solution $\vec{S}(t)$ was then averaged in the upper state with the corresponding lifetime distribution $\tau_{upper}^{-1} \exp(-t/\tau_{upper})$ and transformed into the starting condition for the lower state evolution. Analogously, the lower-state solution was weighted with $\tau_{lower}^{-1} \exp(-t/\tau_{lower})$ and split into the polarization components of the optical response (each the component is equal to twice the corresponding projection of the mean pseudospin[4]).

We allow for random orientations of the principal axes of the QDs' *lower* (ground) state potential (in-plane 'elongations' of QDs, where the inhomogeneous in-plane strain distribution can add to the QD shape effect[15,16]). Confirmed by various ensemble and single-QD experiments, the directional scatter is a well-established reality for many epitaxial QD systems. We shall describe it by a probability density function[8]

$$R(\psi) = \frac{1}{2\pi}[1 + \alpha \cos 2\psi + \beta \cos 4\psi], \tag{2}$$

$\alpha = a/(1+a+b)$, $\beta = b/(1+a+b)$; $a,b > 0$, where $R(\psi)d\psi$ characterizes the fraction of QDs elongated within the $d\psi$ angular range and the direction $\psi = 0$ corresponds to the [110] axis. Here $\alpha$ measures the excessive likelihood to find a QD with the 'long' axis parallel to [110] (over [1$\bar{1}$0], i.e., an overall orthorhombic distortion of the layer) while $\beta$ measures the similar preference of {110}-directions (over {100}-directions, i.e., a reflection of the cubic symmetry of the host lattice). Based on the previous experience,[8,17] we believe that the function Eq.(2) correctly reproduces the main symmetry features of the real directional density functions for the popular QD systems. The directions of the *upper* state potential will be specified in what follows.

Let us establish a convenient notation allowing a compact presentation of results. We shall describe the external and internal parameters and their essential combination in terms of four *phase gains*

$$\Theta_e = \frac{\delta_{1,upper}}{\hbar}\tau_{upper}, \quad \Theta_B = \frac{\mu_B g_{upper} B}{\hbar}\tau_{upper}; \quad \theta_e = \frac{\delta_{1,lower}}{\hbar}\tau_{lower}, \quad \theta_B = \frac{\mu_B g_{lower} B}{\hbar}\tau_{lower}, \qquad (3)$$

where capital $\Theta$s refer to the upper, small $\theta$s – to the lower state. Together with the parameters $\alpha, \beta$ of the angular function Eq.(2) they form the full parameter system of the problem.

We searched for all the polarization components of the luminescence for the cases of linearly ($L$) and circularly ($C$) polarized excitation as functions of the magnetic field $B$ and the angle $\varphi$ the [110] axis of the sample makes with a fixed direction $L$ in the laboratory reference frame (conventionally, 'vertical'). The resulting polarization degrees can be written in a unified form

$$P_{ij} = \frac{1}{1+\Theta_e^2+\Theta_B^2} \cdot \frac{1}{1+\theta_e^2+\theta_B^2} \cdot \left[C_{ij}^{inv} + C_{ij}^{\cos 2\varphi}\cos 2\varphi + C_{ij}^{\sin 2\varphi}\sin 2\varphi + C_{ij}^{\cos 4\varphi}\cos 4\varphi + C_{ij}^{\sin 4\varphi}\sin 4\varphi\right]$$

$$i = L, C; \; j = L, L', C, \qquad (4)$$

where $i$ specifies the incoming-, $j$ – the outgoing polarization, $L'$ stands for the linear polarization degree in the axes rotated by 45° with respect to the vertical direction, coefficients $C_{ij}$ are functions of $B$ but not of $\varphi$.

Specific expressions of the coefficients $C_{ij}$ depend on the conventions regarding the upper exciton state, and here diverge three models that we shall consider. First, a fully correlated conversion, i.e., the excited-state QD potential is elongated in the same direction as the ground-state one. (This is as the two-step model of Ref.8, but here we consider a more general case with no limitations imposed on the ground state lifetime.) The corresponding cells in the Table are marked (corr). By this model we have in mind large QDs (each of them contains several exciton levels) being excited by photons whose energy exceed the PL energy only slightly. Second, a fully non-correlated conversion (marked n/corr), i.e., the excited-state potential has arbitrary direction of the in-plane elongation with no relationship to the ground-state elongation. Third, an almost non-correlated conversion (marked n/corr*), i.e., the excited-state potential is elongated parallel to [110] for all the QDs with no relationship to the ground-state elongation. The latter two models can be compared to the case of small QDs as emitting states and the excitons being generated in the wetting layer. For the n/corr model, the localization of the excited states can be, e.g., on some conglomerates of QDs or 'islands' in a strongly corrupted wetting layer, while for the n/corr* model – in a more perfect quasi-2D layer, a quantum well with non-equivalent interfaces.

**Discussion**

Generally, in the framework of the two-step models of the conversions effect, the polarization response of the system is determined by Eq.(4) and nine non-equivalent field-dependent coefficients $C_{ij}$ (see the Table). Five of these nine are even, other four – odd functions of the magnetic field. Each the coefficient corresponds to some contribution to the PL polarization which can be experimentally measured by choosing an appropriate polarization configuration and by separation of a certain harmonic of the $\varphi$-dependence of the polarization signal.

In fact, the nine coefficients $C_{ij}$ form a very natural basis for classification of a manifold of the conversions effects. We shall see that various experimental manifestations reported by different authors for different systems can be well inscribed in this classification scheme. So perhaps the classification based of the coefficients $C_{ij}$ is a central message of this paper.

We begin the analysis of the Table by a checkup of simple passages to the limit. First, if the anisotropic exchange in the upper stage is small enough (the phase $\Theta_e \to 0$, while $\Theta_B$ is arbitrary), the direction of elongation of the upper-stage potential should be insignificant. Then different versions of the 2-step model must produce the same results, and this is obeyed. If, further, the upper stage is insignificant at all, so that $\Theta_B \to 0$ too (e.g., because the lifetime of the upper state is very short), then all the 2-step scenarios must be reduced to a simple 1-step evolution in the lower state. This is also obeyed (cf. with '1-step' cells in the Table).

Using Eq.(4) and the Table, one can directly calculate the field– and angular dependences of the polarization responses within every scenario and at any parameter values. However, in estimation of typical trends, a further simplification of formulas can be productive. For example, a realistic assumption of long lifetime in the lower stage[8] can be accepted for several existing QD systems. This assumption implies $\theta_e \gg 1$ and $\theta_e \gg \Theta_e$, while the ratio between $\Theta_e$ and unity is not specified since both the upper-state anisotropic exchange and the upper-state lifetime can be, in principle, very different. We shall perform the further analysis with an eye on the above two 'standard' assumptions.

Turning to the one-by-one discussion of the coefficients $C_{ij}$, we start with the coefficient #9. It describes the $CC$ response and is readily identified as 'optical orientation of excitons'. If both exciton states make no conversion of the polarization (i.e., with all the phases $\Theta_e, \Theta_B, \theta_e, \theta_B \to 0$), the initial value of $P_{CC} = 1$ is maintained. Let for a while the external magnetic field is zero ($\Theta_B, \theta_B = 0$). Then the zero-field optical orientation degree $P_{CC}(0)$ is controlled by the values $\Theta_e, \theta_e$ and, under the "standard" assumptions, is small in $\theta_e^{-1}$. This is quite typical for QDs that the optical orientation signal is absent or nearly absent at zero field. But the optical orientation can be *restored* by application of the longitudinal magnetic field; formally it is reproduced by the limit $P_{CC} = 1$ at $\Theta_B, \theta_B \to \infty$ in all the models. Finally, worth commenting is possible inversion of sign of $P_{CC}$ within the (corr) model at zero magnetic field, when $\Theta_e \theta_e > 1$. This behavior is well known from the two-level ('cascade') Hanle effect,[3] with the effective field of the anisotropic exchange interaction acting for the transverse magnetic field of the Hanle effect experiment. Qualitatively, it is quite clear why it is not possible with the non-correlated directions of precession ((n/corr) scheme).

Coefficient #1 in the Table ($C_{LL}^{inv}$) describes the isotropic part of 'optical alignment of excitons' which can be suppressed by the longitudinal magnetic field, the effect sometimes referred to as the longitudinal Hanle effect. This contribution dominated in the experimental $LL$ response of CdSe/ZnSe QDs.[8] To illustrate the distinction of different two-step scenarios, we note that within the (corr) model, at a zero magnetic field, under "standard" assumptions and at $\Theta_e \gg 1$ the maximum value of polarization is ½. This is because the transverse components of the mean pseudospin vanish as a result of long-lasting precession in the chaotically oriented exchange fields. At all the same conditions, the maximum polarization within the (n/corr) model equals ¼ since the transverse components vanish twice – in the upper and in the lower stages.

Regarding coefficient #2, the corresponding contributions to the $LL$ and $LL'$ responses ('swings of optical alignment') were experimentally observed in Ref.8. By its $B$-dependence this coefficient is similar to #1. In majority of schemes this coefficient is proportional to the $\beta$-factor which describes the 90-degrees periodicity in the distribution of elongations (Eq.(2)), i.e., preferential elongations parallel to {110}-type axes. However, within the (n/corr*) scheme it also includes terms which do not contain $\beta$. Such terms appear due to combination of two factors: (i) presence of the fourth angular harmonic in the polarization response of a single QD and (ii) presence of the regular direction of elongation in the upper stage of the (n/corr*) scheme.

Coefficient #3 also concerns with the fourth angular harmonic of the polarization response, but characterizes its $B$-odd component phase-shifted by quarter period. What catches the eye, this coefficient strictly equals zero within the 1-step approach. So once the corresponding polarization is observed, the 1-step scheme is not adequate to the experimental conversions for sure. By its $B$-dependence, #3 resembles coefficient #4, but is smaller in value. (The related two conversions can be conventionally entitled 'anisotropic and isotropic linear-to-linear conversions', respectively). A $B$-odd behavior the $LL'$ response was reported for the first time in Ref.2, but because the angular scan of the effect was missing there, one can't know whether that result should be associated with #3 or #4. Thus a clear observation of the harmonic related to coefficient #3 is a minor challenge for future studies.

Contributions associated with other two $B$-odd coefficients, #5 and #7, can be named 'odd (or conventional) linear-to-circular and circular-to-linear conversions'. It is these conversions that were discovered in the early papers by Blackwood et al.[1] and Dzhioev et al.[2] We note that difference in amplitude between the experimental linear-to-circular and circular-to-linear conversions was reported in Ref.2 and was associated with the two-step evolution of the exciton. Indeed, one can see that while the 1-step model predicts equal amplitudes for the two conversions, any version of 2-step model allows them to be not equal.

Finally, coefficients #6 and #8 describe, in particular, the zero-field linear-to-circular and circular-to-linear conversions reported by Astakhov et al.[6] Unlike the previous pair, coefficients #6 and #8 are even in $B$. According to our classification, this unusual parity in $B$ as well as the specific phase of the second angular harmonic are really distinctive features of the two conversions (rather than just their nonzero value at zero field[6]), so the proper entitlement might be 'even linear-to-circular and circular-to-linear conversions'. In fact, the experimental investigation of the field dependences of the even conversions is of interest. The Figure gives examples of the calculated field dependences of the 'even' conversions. The W- (M-) shaped dependences like those shown in Fig.(a) are not known from experiment, which constitutes a second minor challenge of the present paper.

One can see that at $B = 0$, the amplitudes of the $LC$ and $CL$ conversions are equal within the 1-step model; more interestingly, they are also equal within the (corr) version of the 2-step model. Within the (n/corr) model, the zero-field conversions amplitudes ratio depends on $\Theta_e$: The amplitudes are equal again at $\Theta_e \ll 1$ while the $LC$ conversion exceeds the $CL$ one by $\Theta_e^2/2$ times at $\Theta_e \gg 1$. With all that, both conversions remain small in $\theta_e^{-1}$ under the "standard" assumption $\theta_e \gg 1$. The situation is different within the (n/corr*) model. As compared to the $LC$ conversion (coefficient #6), the $CL$ conversion (coefficient #8) additionally includes the term $\propto \Theta_e \theta_e^2$ which will dominate at a "standard" $\theta_e \gg 1$. Because of this term, the $CL$ conversion is no longer small $\theta_e^{-1}$ and can be as large as 25% if the optimal value $\Theta_e = 1$ is assumed. In effect, the $CL$ conversion itself occurs in the upper short-lifetime state. The resulting circular polarization is then merely "stored" in the lower-level state by means of the optical alignment $LL$, the effect which does not vanish in the limit of long times of the pseudospin precession (coefficients #1 and #2, 1-step model). We note that the $LC$ conversion is still small in $\theta_e^{-1}$. Thus the loss of correlation between the upper and lower level according to the (n/corr*) scenario, even occurring for a fraction of excitons only, can be the reason for the amplitudes ratio $P_{CL} > P_{LC}$ reported from the zero-field experiments in Ref.6.

**Conclusions**

In conclusion, we have calculated the excitonic polarization conversions by a model system of semiconductor quantum dots. The unified presentation of all kinds of the polarization response in the form of Eq.(4) shows relationships between them and gives a natural basis for the classification of the conversions. The classification is based on the 9 non-zero coefficients $C_{ij}$ whose explicit expressions were calculated for several scenarios and presented in the Table.

Overall, the classification scheme well inscribes the results obtained by different authors, including the non-equal values of the zero-field *LC* and *CL* conversions observed in Ref.6. One can see, however, a few lacunas in the well-developed experimental picture of the phenomenon, like the anisotropic linear-to-linear conversion (coefficient #3) and the *B*-even magnetic field pattern of the linear-to-circular and circular-to-linear conversions (coefficients #6 and #8, see the Figure). In order to promote filling these lacunas, we briefly consider (in the Appendix) the symmetry features of sideways contributions to the polarization signals that can be expected to appear for real quantum dot samples.

## Acknowledgments

We are indebted to G.V. Astakhov and T. Kiessling for stimulating discussions, to K.V. Kavokin for valuable comments on the draft. This work was supported by Deutsche Forschungsgemeinschaft and by RFBR, project 09-02-00884.

## APPENDIX: Accompanying contributions to the PL polarization

This paper studies the polarization responses of QD excitons, i.e., different components of the PL polarization signal which are induced by the polarization of the excitation light. But there exist other components of the polarization which have nothing to do with the polarized excitation – they would equally well appear if the luminescence were induced by non-polarized light, or even not by light. Since these components of the PL polarization are constantly observed in experiments, their properties should be clearly recognized and they should be separated for a correct measurement of the polarization responses.

The first accompanying contribution is usually referred to as MCPL – magnetic field induced circular polarization of the PL. It appears, microscopically, due to preferential Boltzmann population of the low-energy bright exciton state, inasmuch this state acquired the helical symmetry in the applied field *B* (it is linearly polarized at $B = 0$). This polarization is not transformed under the sample rotation, thus giving a $\varphi$-independent contribution into the configurations with outgoing *C* light:

$$P_{LC}, P_{CC} = \Re^{inv}(B).$$

This contribution is distinct in the $LC$ configuration since the true conversions do not include any $\varphi$-independent terms in $LC$. It is not so easy with optical orientation $CC$ which is $\varphi$-independent too (coefficient #9). The MCPL and coefficient #9 can be separated using their opposite parity in the magnetic field, since the MCPL is odd in $B$. The more fundamental approach is a separate measurement of $\Re^{inv}(B)$ using the $LC$ configuration or the $NC$ configuration[8] (where $N$ symbolizes the 'non-polarized' state of the excitation light prepared with the polarization scrambler).

The second accompanying contribution is often referred to as built-in linear polarization of the PL. Microscopically, it can be produced by two different mechanisms: (i) preferential Boltzmann population of one bright exciton state like in the MCPL, but inasmuch the state keeps its linear symmetry (temperature-dependent mechanism) or (ii) deformation-induced heavy-light hole mixing (temperature-independent mechanism). Phenomenologically, both lead to a certain amount of linear polarization in the PL, with the direction being along (or perpendicular to) the [110] axis, and with the degree $\wp$ being independent of the type of the incident polarization.

The built-in polarization will manifest itself as a second harmonic of the $\varphi$-dependences of the polarization, a trivial result of the rotation of the sample. It will accompany conversions in the configurations with the outgoing $L$ ($LL$ and $CL$) as

$$P_{LL}, P_{CL} = \wp \cos 2\varphi$$

and will appear in the configurations with outgoing $L'$ as

$$P_{LL'}, P_{CL'} = \wp \sin 2\varphi.$$

Conversions in the $LL$ and $LL'$ configurations do not contain any second harmonic contribution, so it should not be a problem to select the built-in polarization here. The circular-to-linear configurations are more vulnerable, since the coefficient #7 shows absolutely the same angular pattern. Thus the built-in polarization ought be evaluated using the $LL$ or $LL'$ results or, the most reliable, measured independently using the polarization scrambler in the laser beam.[8]

**Table.** Non-zero coefficients $C_{ij}$ for different models as described in the text.

| (Number) Coefficient | Model | Expression |
|---|---|---|
| (1) $C_{LL}^{inv}$ | 2-step (corr) | $1 + \frac{1}{2}\Theta_e^2 + \frac{1}{2}\theta_e^2 - \frac{1}{2}\Theta_e\theta_e - \Theta_B\theta_B + \frac{1}{2}\Theta_e^2\theta_e^2 + \frac{1}{2}\Theta_e\Theta_B\theta_e\theta_B$ |
| | 2-step (n/corr) | $1 + \frac{1}{2}\Theta_e^2 + \frac{1}{2}\theta_e^2 - \Theta_B\theta_B + \frac{1}{4}\Theta_e^2\theta_e^2$ |
| | 2-step (n/corr*) | $1 + \frac{1}{2}\Theta_e^2 + \frac{1}{2}\theta_e^2 - \Theta_B\theta_B + \frac{2+\beta}{8}\Theta_e^2\theta_e^2 - \frac{1}{4}\alpha(\Theta_e\theta_e - \Theta_e\Theta_B\theta_e\theta_B)$ |
| | 1-step | $1 + \frac{1}{2}\theta_e^2$ |
| (2) $C_{LL}^{\cos 4\varphi} = C_{LL'}^{\sin 4\varphi}$ | 2-step (corr) | $\frac{1}{4}\beta(\Theta_e^2 + \Theta_e\theta_e + \theta_e^2 + \Theta_e^2\theta_e^2 + \Theta_e\Theta_B\theta_e\theta_B)$ |
| | 2-step (n/corr) | $\frac{1}{4}\beta\left(\theta_e^2 + \frac{1}{2}\Theta_e^2\theta_e^2\right)$ |
| | 2-step (n/corr*) | $\frac{1}{4}\left(2\Theta_e^2 + \frac{2+\beta}{2}\Theta_e^2\theta_e^2 + \beta\theta_e^2 + \alpha(\Theta_e\theta_e + \Theta_e\Theta_B\theta_e\theta_B)\right)$ |
| | 1-step | $\frac{1}{4}\beta\theta_e^2$ |
| (3) $C_{LL}^{\sin 4\varphi} = -C_{LL'}^{\cos 4\varphi}$ | 2-step (corr) | $\frac{1}{4}\beta(\Theta_B\theta_e^2 - \Theta_e^2\theta_B + \Theta_e\Theta_B\theta_e - \Theta_e\theta_e\theta_B)$ |
| | 2-step (n/corr) | $\frac{1}{4}\beta\Theta_B\theta_e^2$ |
| | 2-step (n/corr*) | $\frac{1}{4}(-2\Theta_e^2\theta_B + \beta\Theta_B\theta_e^2 + \alpha(\Theta_e\Theta_B\theta_e - \Theta_e\theta_e\theta_B))$ |
| | 1-step | $0$ |
| (4) $C_{LL'}^{inv}$ | 2-step (corr) | $\Theta_B + \theta_B + \frac{1}{2}(\Theta_e^2\theta_B + \Theta_B\theta_e^2 - \Theta_e\Theta_B\theta_e - \Theta_e\theta_e\theta_B)$ |
| | 2-step (n/corr) | $\Theta_B + \theta_B + \frac{1}{2}(\Theta_e^2\theta_B + \Theta_B\theta_e^2)$ |
| | 2-step (n/corr*) | $\Theta_B + \theta_B + \frac{1}{2}(\Theta_e^2\theta_B + \Theta_B\theta_e^2) - \frac{1}{4}\alpha(\Theta_e\Theta_B\theta_e + \Theta_e\theta_e\theta_B)$ |
| | 1-step | $\theta_B$ |
| (5) $C_{LC}^{\cos 2\varphi}$ | 2-step (corr) | $\frac{1}{2}\alpha(\Theta_e\Theta_B + \Theta_B\theta_e + \theta_e\theta_B + \Theta_e^2\theta_e\theta_B + \Theta_e\Theta_B\theta_B^2)$ |
| | 2-step (n/corr) | $\frac{1}{2}\alpha\left(\Theta_B\theta_e + \theta_e\theta_B + \frac{1}{2}\Theta_e^2\theta_e\theta_B\right)$ |
| | 2-step (n/corr*) | $\Theta_e\Theta_B + \Theta_e\Theta_B\theta_B^2 + \frac{1}{2}\alpha(\Theta_B\theta_e + \theta_e\theta_B + \Theta_e^2\theta_e\theta_B)$ |
| | 1-step | $\frac{1}{2}\alpha\theta_e\theta_B$ |
| (6) $C_{LC}^{\sin 2\varphi}$ | 2-step (corr) | $-\dfrac{1}{2}\alpha\left(\Theta_e + \theta_e + \Theta_e\theta_B^2 - \Theta_B\theta_e\theta_B\right)$ |
| | 2-step (n/corr) | $-\dfrac{1}{2}\alpha\left(\theta_e + \dfrac{1}{2}\Theta_e^2\theta_e - \Theta_B\theta_e\theta_B\right)$ |
| | 2-step (n/corr*) | $-\Theta_e - \Theta_e\theta_B^2 - \dfrac{1}{2}\alpha(\theta_e - \Theta_B\theta_e\theta_B)$ |
| | 1-step | $-\dfrac{1}{2}\alpha\theta_e$ |
| (7) $C_{CL}^{\cos 2\varphi} = C_{CL'}^{\sin 2\varphi}$ | 2-step (corr) | $\dfrac{1}{2}\alpha\left(\Theta_e\Theta_B + \Theta_e\theta_B + \theta_e\theta_B + \Theta_e\Theta_B\theta_e^2 + \Theta_B^2\theta_e\theta_B\right)$ |
| | 2-step (n/corr) | $\dfrac{1}{2}\alpha\left(\theta_e\theta_B + \Theta_B^2\theta_e\theta_B\right)$ |
| | 2-step (n/corr*) | $\Theta_e\Theta_B + \Theta_e\theta_B + \dfrac{2+\beta}{4}\Theta_e\Theta_B\theta_e^2 + \dfrac{1}{2}\alpha(\theta_e\theta_B + \Theta_B^2\theta_e\theta_B)$ |
| | 1-step | $\dfrac{1}{2}\alpha\theta_e\theta_B$ |
| (8) $C_{CL}^{\sin 2\varphi} = -C_{CL'}^{\cos 2\varphi}$ | 2-step (corr) | $\dfrac{1}{2}\alpha\left(\Theta_e + \theta_e + \Theta_B^2\theta_e - \Theta_e\Theta_B\theta_B\right)$ |
| | 2-step (n/corr) | $\dfrac{1}{2}\alpha\left(\theta_e + \Theta_B^2\theta_e\right)$ |
| | 2-step (n/corr*) | $\Theta_e - \Theta_e\Theta_B\theta_B + \dfrac{2-\beta}{4}\Theta_e\theta_e^2 + \dfrac{1}{2}\alpha(\theta_e + \Theta_B^2\theta_e)$ |
| | 1-step | $\dfrac{1}{2}\alpha\theta_e$ |
| (9) $C_{CC}^{inv}$ | 2-step (corr) | $1 + \Theta_B^2 + \theta_B^2 - \Theta_e\theta_e + \Theta_B^2\theta_B^2 + \Theta_e\Theta_B\theta_e\theta_B$ |
| | 2-step (n/corr) | $1 + \Theta_B^2 + \theta_B^2 + \Theta_B^2\theta_B^2$ |
| | 2-step (n/corr*) | $1 + \Theta_B^2 + \theta_B^2 + \Theta_B^2\theta_B^2 - \dfrac{1}{2}\alpha(\Theta_e\theta_e - \Theta_e\Theta_B\theta_e\theta_B)$ |
| | 1-step | $1 + \theta_B^2$ |

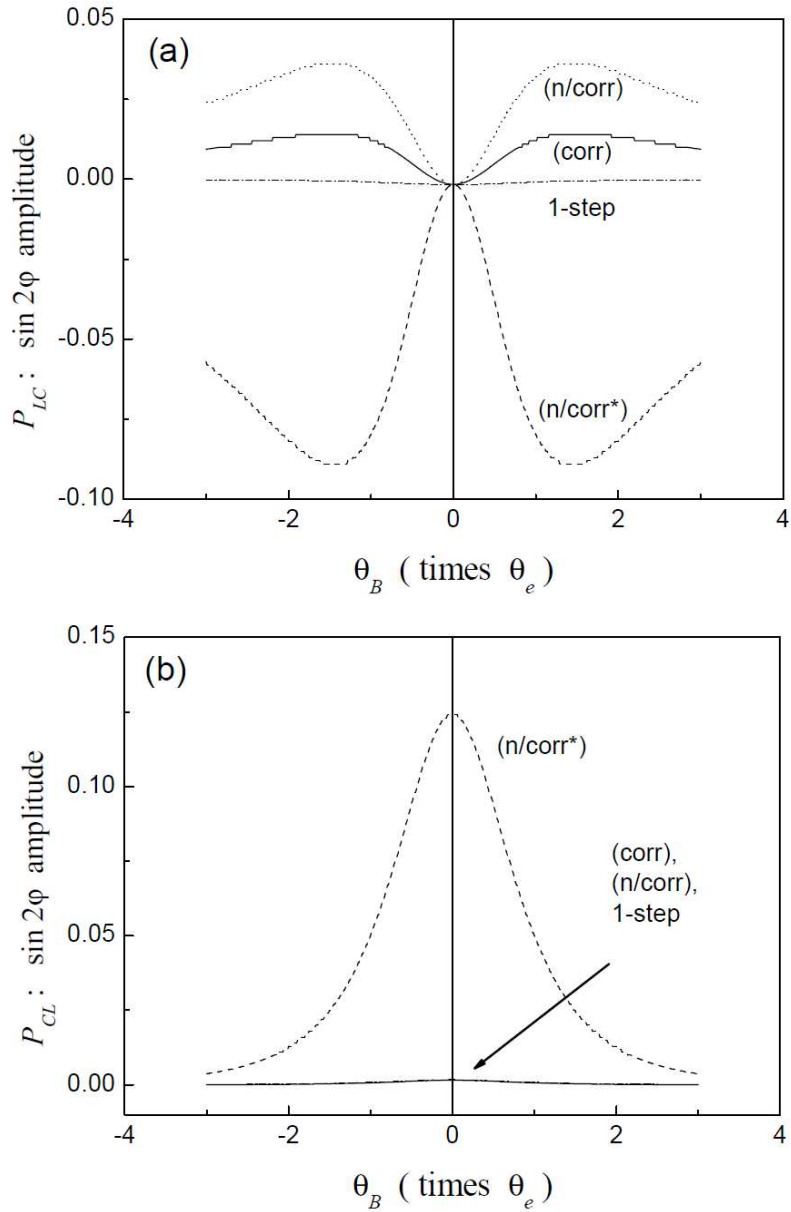

**Figure**. 'Even' linear-to-circular (a) and circular-to-linear (b) polarization conversions calculated for different scenarios of evolution, as functions of the magnetic field $B$ (the latter is quantified by the dimensionless phase gain $\theta_B$ which is shown, however, in units of $\theta_e$). A realistic sample parameter set was used: $\theta_e = 100$ ($\delta_{1,lower} \sim 0.3$ meV, $\tau_{lower} \sim 200$ ps), $\Theta_e = 0.3$, $\theta_B/\Theta_B = 200$, and $\alpha = 0.35$, $\beta = 0.22$ as obtained in Ref.8 ($a = 0.8$, $b = 0.5$). Calculated with Eq.(4) and coefficients #6 and #8 from the Table.